\documentclass[draft,grl]{agutex}

  \usepackage[dvips]{graphicx}
  \setkeys{Gin}{draft=false}
\authorrunninghead{YERMOLAEV}

\titlerunninghead{DURATIONS OF STORM PHASES}

\begin{document}

%
%

\title{Whether duration of the recovery phase of magnetic storm depends on the development rate of storm at its main phase?}
%
%

%
%









\authors{Yuri I. Yermolaev\altaffilmark{1}}

\altaffiltext{1}{Space Plasma Physics Department, Space Research Institute, 
Russian Academy of Sciences, Profsoyuznaya 84/32, Moscow 117997, Russia. 
(yermol@iki.rssi.ru)}

%
%


\begin{abstract}
We compare dependences between the storm development rate $|Dst_{min}|/\Delta T$ 
($\Delta T$ is the durations of main phase) and the duration of recovery phase of magnetic storms 
generated by three various types of interplanetary drivers: 
(1, 2) compression regions CIR and Sheath, and 
(3) body of interplanetary CME (magnetic clouds and Ejecta). 
Our analyze shows that the duration of recovery phase correlates with the storm development rate 
for CIR- and Sheath-induced storms, and does not correlate for ICME- induced storms. 
\end{abstract}

%
%

%

\begin{article}

%
%

\section{Introduction}
As has been shown by numerous papers,  
the dynamics of the magnetosphere during the development of magnetic storms 
significantly depends on the type of interplanetary driver
(see, e.g.,  
\citep{Huttunenetal2002,Huttunenetal2006,BorovskyDenton2006,Pulkkinenetal2007,Yermolaevetal2010,Yermolaevetal2012a,Guoetal2011,LiemohnKatus2012,Nikolaevaetal2013,Crameretal2013} 
and references therein).  
These types of drivers are following: body of interplanetary CME (ICME) 
including magnetic cloud (MC) and Ejecta, and compression regions 
before high-speed solar wind stream (corotating interplanetary region -- CIR) 
and before ICME (Sheath)
(see, e.g., 
\citep{Gosling1993,Gonzalezetal1999,Yermolaevetal2005}).  
Recently we showed that the dynamics of recovery phase of magnetic storms depends 
on the interplanetary driver types 
\citep{Yermolaevetal2014}. 
In particular we found that the durations of the main phase and the recovery phase correlate 
for CIR- and Sheath-induced storms and there is no dependence for these durations 
for MC- and Ejecta-induced storms. In this brief paper we will analyze data in details and  
give a physical interpretation of this result. 

\section{Methods}
We use the same data set that we used in our previous papers
\citep{Yermolaevetal2012,Yermolaevetal2014}: 
the measurements of $Dst$ index (see http://wdc.kugi.kyoto-u.ac.jp/index.html) 
and our catalog of large-scale interplanetary events for the period of 
1976--2000 (see the web site ftp://www.iki.rssi.ru/pub/omni and paper by 
\cite{Yermolaevetal2009}),
prepared on the basis of OMNI dataset of interplanetary plasma and magnetic field parameters 
(see http://omniweb.gsfc.nasa.gov) and paper by  
\cite{KingPapitashvili2004}).
The detailed description of the technique of the solar wind classification and 
comparison to magnetic storms is provided in several papers
\citep{Yermolaevetal2009,Yermolaevetal2010,Yermolaevetal2012a,Yermolaevetal2012}.

Method of determination of durations of the main and recovery phases is schematically shown 
in Figure 1. 
To consider the existence of fast (initial) and slow (second) parts of the recovery
(see, e.g., 
\citep{Gonzalezetal1994}), we calculate two durations: 
the initial time interval from the minimum of the $Dst$ index up to $(1/2)Dst_{min}$ 
($\Delta t_{1/2} = t((1/2)Dst_{min}) - t(Dst_{min})$) and $(1/3)Dst_{min}$ 
($\Delta t_{1/3} = t((1/3)Dst_{min}) - t(Dst_{min})$), respectively. 
Analysis of the two durations  
$\Delta t_{1/2}$ and $\Delta t_{1/3}$ allows us to 
compare the durations of the fast and slow parts of the recovery phase.

\section{Results}

In our previous paper 
\citep{Yermolaevetal2014}
we studied the durations of main $\Delta T$ and recovery 
$\Delta t_{1/2}$ and $\Delta t_{1/3}$ phases of storms induced 
by different interplanetary drivers and found the anticorrelation for CIR- and Sheath-induced storms. 
It is naturally to suggest that all durations depend on the magnitude of storms. 
However, the selection of data on the storm magnitude decreases the number of events 
and accuracy of analyze, and did not allow us to make reliable conclusions.  
So, we study a new variable 
$|Dst_{min}|/\Delta T$
which includes both duration $\Delta T$  and storm magnitude $Dst_{min}$, 
and is an average temporal derivation of $Dst$ index or a storm development rate. 
 
Figure 2 presents the dependence between the storm development rate $Dst_{min}/\Delta T$ and 
the fast and slow durations of recovery phase ($\Delta t_{1/2}$ and $\Delta t_{1/3}$) for different drivers. 
Three lower panels (from the bottom panel up) show individual events for Sheath-, CIR- and ICME-induced storms, 
left and right coulombs present data for fast and slow durations. 
The straight lines through the data are linear fits to the points in the two-logarithmic scale 
(i.e., power law approximations of data).
The top panels represent results of data fitting for the lower panels and 
allow one to compare them for various drivers. 
The correlation coefficients $r$ for all panels are presented in Table. 
To emphasize the statistical significance of the results, we present the parameter 
$w = 0.5 \sqrt{(N-3)} ln[(1+r)/(1-r)]$ and probability $P$ 
\citep{BendatPiersol1971}.

Sheath-induced storms have the most deep dependence between the storm development rate and 
both fast and slow durations. 
Despite a wide spread of points, these dependences possess rather high statistical significance 
(Probability $P$ = 90 and 95 \%).
CIR-induced storms have similar parameters for fast duration $\Delta t_{1/2}$ 
but for slow durations $\Delta t_{1/3}$ the fitting line inclination decreases 
with simultaneous decreasing correlation coefficient $r$ and probability $P$. 
ICME-induced storms have low values of line inclination, correlation coefficient and 
probability for both types of recovery durations, i.e. there is no dependence 
between the storm development rate and both fast and slow durations. 

\section{Discussion}
In accordance with formula by 
\cite{Burtonetal1975}
in the case of neglecting the term related to the decay of the ring current at the main phase 
and numerous papers (see, e.g., 
\cite{Kane2010,Ontiveros2010,Yermolaevetal2010,Weigel2010,Nikolaevaetal2013,Nikolaevaetal2014}), 
for various interplanetary drivers the measured and density-corrected $Dst$ and $Dst*$ indexes 
may be approximated by a linear function of the integral of 
interplanetary convective electric field $E_y$  
with high accuracy (the correlation coefficients are 0.98–-0.99), 
i.e. derivative of $Dst$ index is proportional to electric field $E_y: d Dst/d t = C E_y$. 
As approximate equality $Dst_{min}/\Delta T \approx d Dst/d t = C E_y$ is fair, 
the obtained results give the indirect indications in favor of a hypothesis 
that the durations of recovery for magnetic storms induced by CIR and Sheath correlate with  
the average electric field during main phase $<E_y>$. 

Reduction of correlation for slow recovery duration $\Delta t_{1/3}$ for CIR-induced storms 
can be explained by the fact that at the second part of recovery phase 
the external factors start prevailing over internal magnetospheric processes, 
and the high-speed solar wind after CIR is characterized by higher level of disturbances of 
plasma and magnetic field parameters (see, e.g., 
\citep{Hajraetal2014}), 
than in the ICMEs after Sheath. 

As we showed earlier 
(\citep{Nikolaevaetal2013}), 
the interplanetary-magnetospheric coupling coefficient between the derivative of $Dst$ index
and average electric field $<E_y>$ depends on the driver type. 
Therefore the lower correlation in Figure 2 
for ICME-induced storms than for CIR/Sheath-induced storms 
can be connected with lower coupling coefficients for MC/Ejecta-induced storms in comparison with 
coupling coefficients for CIR/Sheath-induced storms.

\section{Conclusions} 
We analyzed the temporal profiles of $Dst$ index for magnetic storms 
induced by various types of interplanetary drivers: 
compression regions CIR (85 storms) and Sheath (71), and bodies of interplanetary CME (158). 
In addition to our previous paper 
\citep{Yermolaevetal2014} 
where we compared the durations of main phase $\Delta T$ and 
the short and long durations of recovery phase  $\Delta t_{1/2}$ and  $\Delta t_{1/3}$ 
(respectively, at the levels of $1/2 Dst_{min}$ and $1/3 Dst_{min}$), 
here we study the dependences of the development rate $Dst_{min}/\Delta T$ 
on recovery phase  durations $\Delta t_{1/2}$ and  $\Delta t_{1/3}$. 
Obtained results allows us to make following conclusions. 

1. The storm development rate $Dst_{min}/\Delta T$ correlates with  
both short $\Delta t_{1/2}$ and long $\Delta t_{1/3}$ durations for Sheath-induced storms.
 
2. The storm development rate $Dst_{min}/\Delta T$ correlates with  
only short duration $\Delta t_{1/2}$ and  does not correlate with long $\Delta t_{1/3}$ 
duration for CIR-induced storms. 
The absence of correlation with long duration may be connected with 
high variability of solar wind and IMF parameters in the high-speed streams 
after CIR regions (see, e.g., 
\citep{Hajraetal2014}), 
in contrast with smooth changing parameters in ICME bodies 
after Sheath regions. 

3. These results allow us to suggest that the physical processes of 
formation and decay of storm activity in the magnetosphere are similar 
for CIR- and Sheath-induced storms. 

4. There is no correlation between the storm development rate $Dst_{min}/\Delta T$ and 
short and long durations of recovery phase for ICME-induced storms. 

5. The magnetosphere processes, which are responsible for storm activity, are suggested 
to be different for ICME- and CIR/Sheath-induced storms. This may be connected with various 
interplanetary-magnetospheric coupling coefficients for different interplanetary drivers  
(\citep{Nikolaevaetal2013}).  


%
%
%
%
%
%
%

\begin{acknowledgments}
The author is grateful for the opportunity to use the OMNI database. The OMNI data were obtained from 
GSFC/ SPDF OMNIWeb (http://omniweb.gsfc.nasa.gov). This work was supported by the 
Russian Foundation for Basic Research, project 13--02--00158, and by 
Program 22 of Presidium of the Russian Academy of Sciences.
\end{acknowledgments}

\end{article}
%
%
%
%
%
%
%
%
\begin{figure}
\noindent\includegraphics[width=10cm]{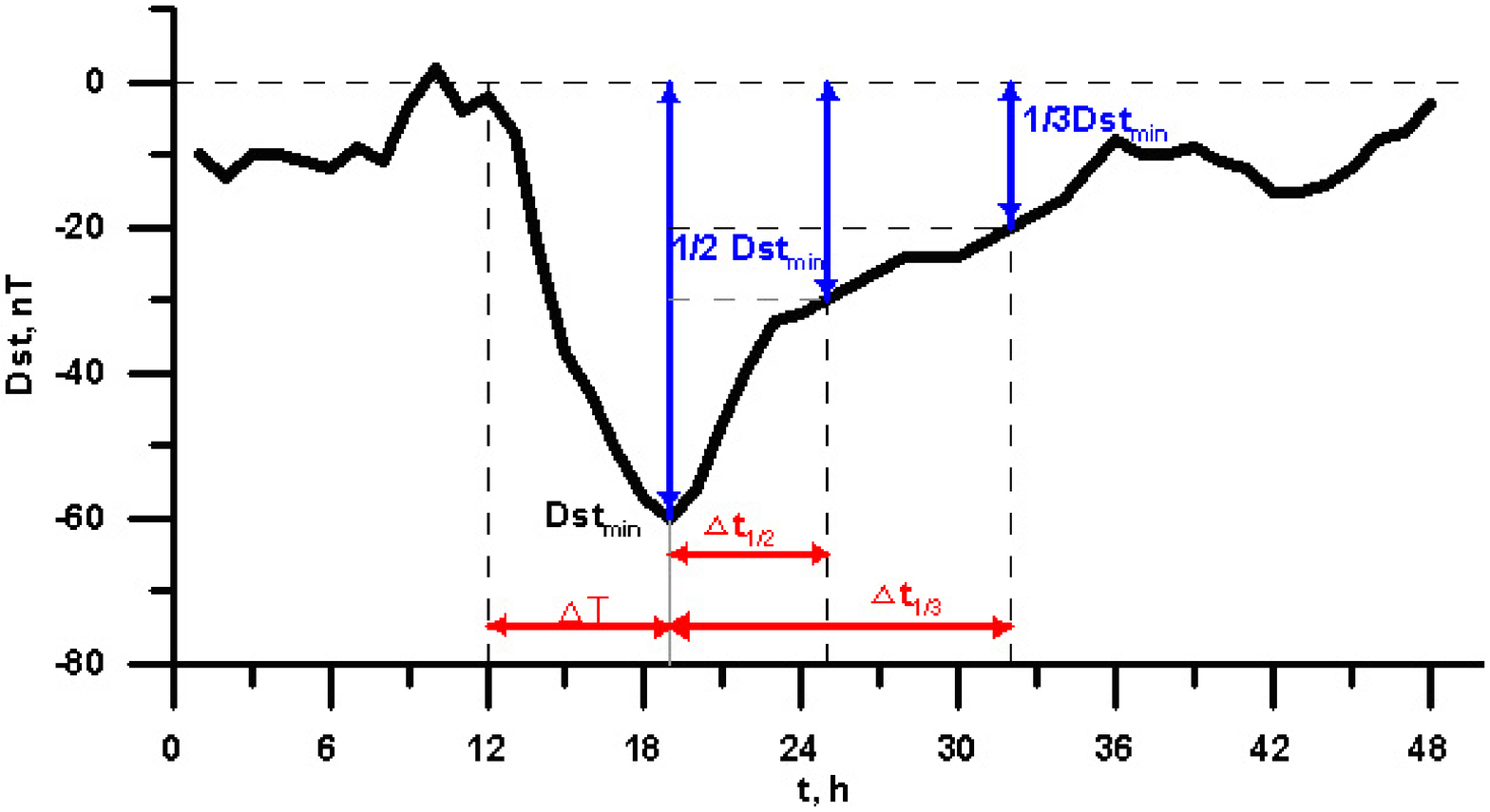}
\caption{Schematic view of method for determination of durations of main and recovery phases for magnetic storms}
\end{figure}

%
\begin{figure}
\noindent\includegraphics[width=10cm]{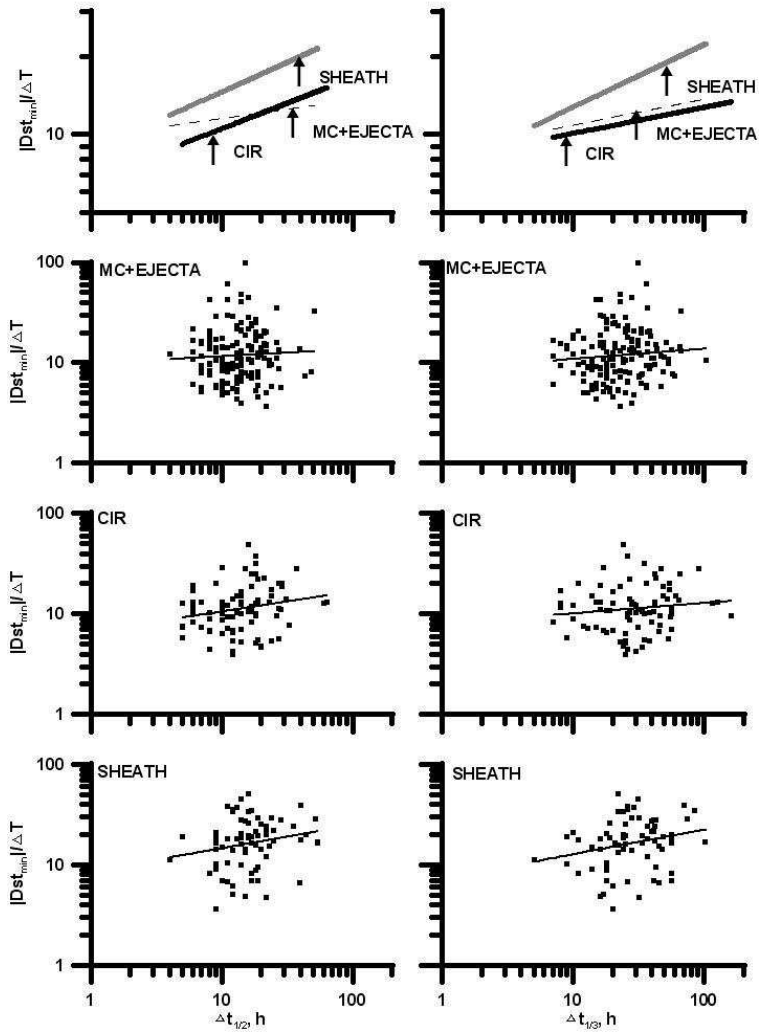}
\caption{Relations between the storm development rate $Dst_{min}/\Delta T$ 
and recovery phase  durations $\Delta t_{1/2}$ and  $\Delta t_{1/3}$
 for magnetic storms generated by CIR, Sheath and ICME (MC+Ejecta)}
\end{figure}
%
%


\begin{table}
{\small 
\caption{The correlation coefficients $r$, probability $P$ and fitting lines 
for relations between the storm development rate $Dst_{min}/\Delta T$ 
and recovery phase  durations $\Delta t_{1/2}$ and  $\Delta t_{1/3}$
 for magnetic storms generated by CIR, Sheath and ICME (MC+Ejecta)
in Figure 2
}
\centering
\begin{tabular}{l|c|cccc|cccc}
\hline
SW & &  \multicolumn{4}{c}{$\Delta t_{1/2}$\tablenotemark{a}} & 
        \multicolumn{4}{|c}{$\Delta t_{1/3}$\tablenotemark{b}} \\
\cline{3-10}        
type   & N & r & P & W & Fitting equation & r & P & W & Fitting equation \\        
\hline
Sheath & 71 & 0.20 & 0.90 & 1.67 & ln y = 0.23 ln x + 2.15 & 0.23 & 0.95 & 1.97 & ln y = 0.24 ln x + 1.99 \\
CIR    & 85 & 0.21 & 0.94 & 1.90 & ln y = 0.20 ln x + 1.90 & 0.15 & 0.73 & 1.10 & ln y = 0.10 ln x + 2.08 \\
ICME\tablenotemark{c} 
      & 158 & 0.05 & 0.49 & 0.66 & ln y = 0.07 ln x + 2.28 & 0.09 & 0.72 & 1.09 & ln y = 0.10 ln x + 2.15 \\
\hline 
\end{tabular}
\tablenotetext{a}{$\Delta t_{1/2} = t(1/2 Dst_{min}) - t(Dst_{min})$}
\tablenotetext{b}{$\Delta t_{1/3} = t(1/3 Dst_{min}) - t(Dst_{min})$}
\tablenotetext{c}{ICME is MC + Ejecta}
}
\end{table}


\begin{thebibliography}{}

\bibitem[{\textit{Bendat and Piersol}(1971)}]{BendatPiersol1971}
Bendat, J. S. and Piersol, A. G., (1971)
Measurement and Analysis of Random Data,  Wiley-Interscience. 

\bibitem[{\textit{Borovsky and Denton}(2006)}]{BorovskyDenton2006}
Borovsky, J. E. and Denton, M.H. (2006), 
Differences between CME-Driven Storms and CIR-Driven Storms,
J. Geophys. Res., 28, 121--190.

\bibitem[{\textit{Burton et al.}(1975)}]{Burtonetal1975}
Burton, R. K., McPherron, R. L., and Russell, C. T (1975), 
An empirical relationship between 
interplanetary conditions and Dst, J. Geophys. Res.,  80, 4204--4214. 

\bibitem[{\textit{Cramer et al.}(2013)}]{Crameretal2013}
Cramer, W. D., N. E. Turner, M.-C. Fok, and N. Y. Buzulukova (2013), 
Effects of different geomagnetic storm drivers on the ring current: CRCM results, 
J. Geophys. Res. Space Physics, 118, doi:10.1002/jgra.50138

\bibitem[{\textit{Gonzalez et al.}(1994)}]{Gonzalezetal1994}
Gonzalez, W. D., Joselyn, J. A., Kamide, Y., Kroehl, H. W., Rostoker, G., Tsurutani, B. T. and
Vasyliunas, V. M. (1994),  What is a Geomagnetic Storm?, J. Geophys. Res. 99, 5771.

\bibitem[{\textit{Gonzalez et al.}(1999)}]{Gonzalezetal1999}
Gonzalez, W. D., Tsurutani, B. T. and Clúa de Gonzalez, A. L. (1999) 
Interplanetary origin of geomagnetic storms, 
Space Science Reviews, Volume 88, Issue 3-4, pp 529-562 

\bibitem[{\textit{Gosling}(1993)}]{Gosling1993}
Gosling, J. T. (1993), The solar flare myth, J. Geophys. Res., 98(A11), 18937–18949, doi:10.1029/93JA01896.

\bibitem[{\textit{Guo et al.}(2011)}]{Guoetal2011}
Guo, J., X. Feng, B. A. Emery, J. Zhang, C. Xiang, F. Shen, and W. Song (2011), 
Energy transfer during intense geomagnetic storms driven by interplanetary coronal mass ejections 
and their sheath regions, J. Geophys. Res., 116, A05106, 
doi:10.1029/2011JA016490 

\bibitem[{\textit{Hajra et al.}(2014)}]{Hajraetal2014}
Hajra, R., E. Echer, B. T. Tsurutani, and W. D. Gonzalez (2014), 
Solar wind-magnetosphere energy coupling efficiency and partitioning: HILDCAAs
and preceding CIR storms during solar cycle 23, J. Geophys. Res. Space Physics,
119, doi:10.1002/2013JA019646.

\bibitem[{\textit{Huttunen et al.}(2002)}]{Huttunenetal2002}
Huttunen K.E.J., Koskinen H.E.J., Schwenn R. (2002), Variability
of magnetospheric storms driven by different solar
wind perturbations, J. Geophys. Res.,  vol. 107, A7, p. 1121.

\bibitem[{\textit{Huttunen et al.}(2006)}]{Huttunenetal2006}
Huttunen, K.E.J., Koskinen, H.E.J., Karinen, A., and Mursula, K. (2006), 
Asymmetric Development of Magnetospheric Storms during Magnetic Clouds and Sheath Regions, 
Geophys. Res. Lett., vol. 33, p. L06107.doi: 10.1029/2005GL024894. 

\bibitem[{\textit{Kane}(2010)}]{Kane2010}
Kane, R.P. (2010), Severe geomagnetic storms and Forbush
decreases: interplanetary relationships reexamined,
Ann. Geophys., 2010, vol. 28, pp. 479--489

\bibitem[{\textit{King and Papitashvili}(2004)}]{KingPapitashvili2004}
King, J.H. and Papitashvili, N.E., (2004), 
Solar Wind Spatial Scales in and Comparisons of Hourly Wind and ACE Plasma and 
Magnetic Field Data, J. Geophys. Res.,vol. 110, no. A2, p. A02209. doi: 10.1029/2004JA010804.

\bibitem[{\textit{Liemohn and Katus}(2012)}]{LiemohnKatus2012}
 Liemohn, M. W., and R. Katus (2012), Is the storm time response of the inner magnetospheric 
hot ions universally similar or driver dependent?, 
J. Geophys. Res., 117, A00L03, doi:10.1029/2011JA017389. 

\bibitem[{\textit{Nikolaeva et al.}(2013)}]{Nikolaevaetal2013}
Nikolaeva, N. S., Yu. I. Yermolaev, and I. G. Lodkina, (2013), 
Modeling of Dst-index temporal profile on the main phase of the magnetic storms 
generated by different types of solar wind, 
Cosmic Research, 2013, Vol. 51, No. 6, pp. 401--412 
(Kosmicheskie Issledovaniya , v. 51, ¹ 6, pp. 443--454) 

\bibitem[{\textit{Nikolaeva et al.}(2014)}]{Nikolaevaetal2014}
Nikolaeva, N. S., Yu. I. Yermolaev, and I. G. Lodkina, (2014), 
Dependence of Geomagnetic Activity during Magnetic Storms on the Solar Wind Parameters 
for Different Types of Streams: 4. Simulation for Magnetic Clouds, 
Geomagnetism and Aeronomy, 2014, Vol. 54, No. 2, pp. 152-161

\bibitem[{\textit{Ontiveros}(2010)}]{Ontiveros2010}
Ontiveros, V. (2010) Geomagnetic storms caused by shocks
and ICMEs, J. Geophys. Res., 2010, vol. 115, A10244.
doi: 10.1029/2010JA015471

\bibitem[{\textit{Pulkkinen et al.}(2007)}]{Pulkkinenetal2007}
Pulkkinen, T. I., Partamies, N., Huttunen, K. E. J., Reeves, G.
D., and Koskinen, H. E. J.: Differences in geomagnetic storms
driven by magnetic clouds and ICME sheath regions, Geophys.
Res. Lett., 34, L02105, doi:10.1029/2006GL027775, 2007

\bibitem[{\textit{Weigel}(2010)}]{Weigel2010}
Weigel, R.S.(2010) Solar wind density influence on geomagnetic storm intensity, 
J. Geophys. Res., 2010, vol. 115,
A09201. doi: 10.1029/2009JA015062.

\bibitem[{\textit{Yermolaev et al.}(2005)}]{Yermolaevetal2005}
Yermolaev, Yu. I.,  Yermolaev, M. Yu.,  Zastenker, G. N., Zelenyi, L. M., 
Petrukovich, A. A., and Sauvaud, J.-A. (2005), 
Statistical studies of geomagnetic storm dependencies on solar and interplanetary events: a review, 
Planetary and Space Science, 
Volume 53, Issues 1–3, Pages 189–196
 
\bibitem[{\textit{Yermolaev et al.}(2009)}]{Yermolaevetal2009} 
Yermolaev, Yu. I., et al., (2009), 
Catalog of Large-Scale Solar Wind Phenomena during 1976--2000, 
Kosm. Issled.,  vol. 47, no. 2, pp. 99--113. [Cosmic Research, pp. 81--94].

\bibitem[{\textit{Yermolaev et al.}(2010)}]{Yermolaevetal2010} 
Yermolaev, Yu. I., N. S. Nikolaeva, I. G. Lodkina, and M. Yu. Yermolaev, (2010), 
Specific interplanetary conditions for CIR-, Sheath-, and ICME-induced geomagnetic storms 
obtained by double superposed epoch analysis, {\it Ann. Geophys.,}  \textit{28}, 2177--2186.

\bibitem[{\textit{Yermolaev et al.}(2012a)}]{Yermolaevetal2012a}
Yermolaev, Y. I., N. S. Nikolaeva, I. G. Lodkina, and M. Y. Yermolaev (2012a), 
Geoeffectiveness and efficiency of CIR, sheath, and ICME in generation of magnetic storms, 
J. Geophys. Res., 117, A00L07, doi:10.1029/2011JA017139

\bibitem[{\textit{Yermolaev et al.}(2012b)}]{Yermolaevetal2012}
Yermolaev, Y. I., I. G. Lodkina, N. S. Nikolaeva, and M. Y. Yermolaev (2012b),
Recovery phase of magnetic storms
induced by different interplanetary drivers, J. Geophys. Res., 117, A08207, doi:10.1029/2012JA017716

\bibitem[{\textit{Yermolaev et al.}(2014)}]{Yermolaevetal2014}
Yermolaev, Y. I., I. G. Lodkina,N. S. Nikolaeva, and M. Y. Yermolaev (2014), 
Influence of the interplanetary driver type on the durations of
the main and recovery phases of magnetic storms, 
J. Geophys. Res. Space Physics, 119, 8126–8136, doi:10.1002/2014JA019826.


%
%
%
%

\end{thebibliography}
\end{document}